\documentclass[5p,sort&compress,number]{elsarticle}

\usepackage{amssymb}
\usepackage{amssymb}
\usepackage{amsmath}

\journal{Nuclear Instruments and Methods in Physics Research Section A}

\begin{document}

\begin{frontmatter}



\title{Calibration systems of the ANTARES neutrino telescope}


\author{J.P. G\'omez-Gonz\'alez on behalf of the ANTARES collaboration}

\address{jpablo@ific.uv.es, IFIC - Instituto de F\'isica Corpuscular, Universitat de Val\`encia-CSIC, E-46015, Valencia, Spain}

\begin{abstract}
The ANTARES detector \cite{bib_antares} is the largest deep sea underwater
neutrino telescope in operation. The apparatus comprises a matrix of 885 photomultiplier tubes (PMTs) which
detect the Cherenkov light emitted by the charged leptons produced in the charged current interactions of
high energy neutrinos with the matter inside or near the detector.
Reconstruction of the muon track and energy can be achieved using the time, position and charge information of the
hits arriving to the PMTs. A good calibration of the detector is necessary in order to ensure its optimal performance.
This contribution reviews the different calibration systems and methods developed by the ANTARES Collaboration.
\end{abstract}


\end{frontmatter}


\section{Introduction}
\label{intro}
Neutrino astronomy opens a new window to explore the most extreme regions of the Universe.
The advantage of using neutrinos as cosmic probes is that, as weakly interacting and almost massless particles, they can point back to the sources of emission without being deflected
or absorbed on their trip. On the other hand, because of very low interaction cross section of the neutrino large instrumented volumes are required to
detect a few events in a year. The ANTARES detector is the largest underwater neutrino telescope in the world.
The apparatus operates collecting the Cherenkov photons emitted by the secondary leptons emerging from 
the interaction of neutrinos by means of a three-dimensional array of light sensors.
The main scientific goal of the experiment is the detection of neutrinos of astrophysical origin.
To this end, the angular resolution is the key parameter. 
Detailed simulations have shown that
for ANTARES it can be as good as 0.3 degrees for neutrino energies above 10 TeV.
In this energy range the angular resolution is dominated by the charged track reconstruction accuracy.
Since the reconstruction algorithms used are based on the time, position and charge information of the signals recorded by the PMTs, a good
calibration is required to ensure an optimal performance of the telescope.

\section{The ANTARES detector}
\label{antares}
\begin{figure}
\begin{center}
\includegraphics[width=0.8\linewidth]{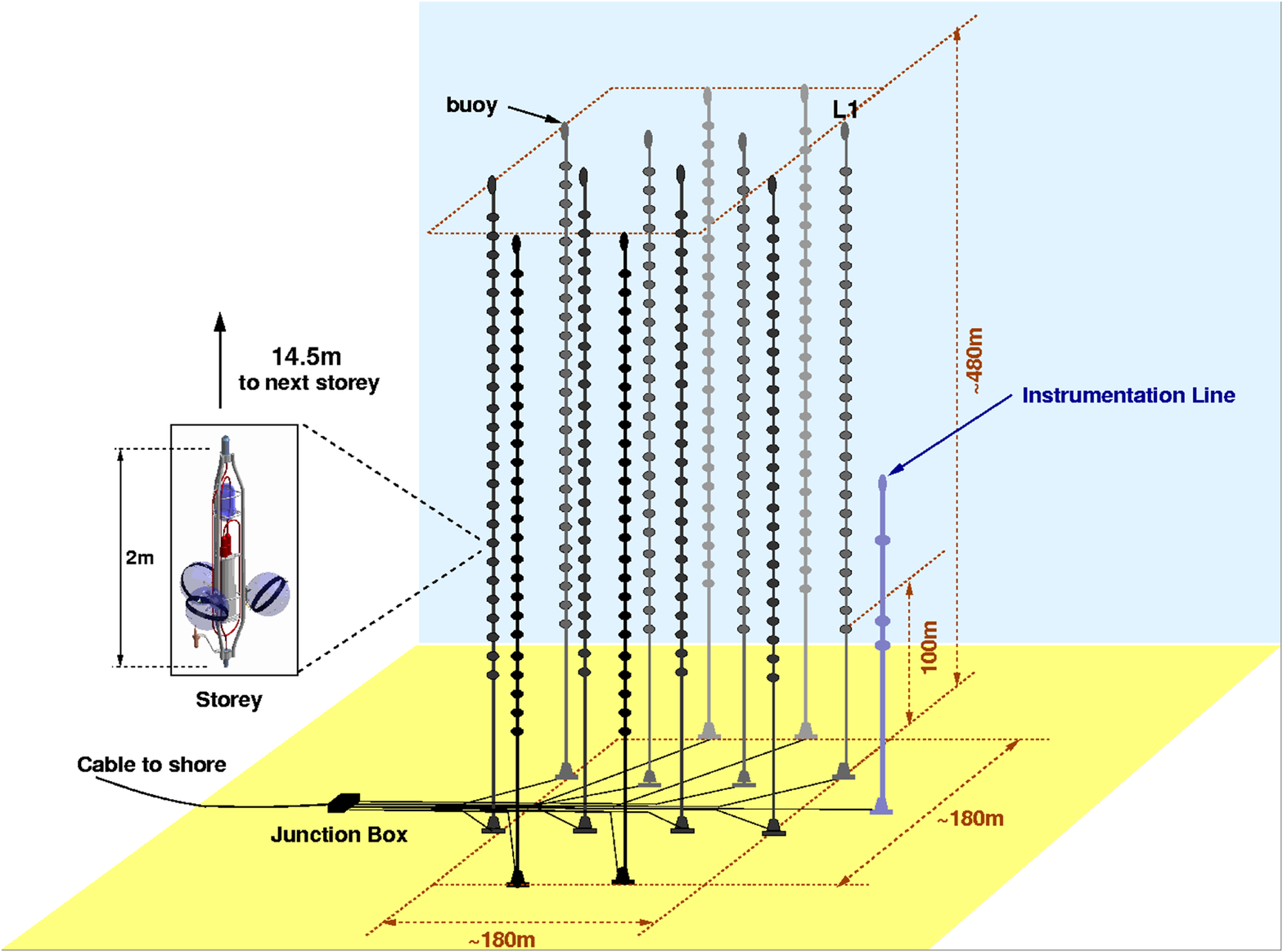}
\caption{Schematic view of the ANTARES detector composed by twelve lines holding triplets of PMTs arranged in 25 storeys.}
\label{fig_antares}
\end{center}
\end{figure}

The ANTARES neutrino telescope is located at a depth of 2475 m 
in the Mediterranean Sea ($42^{\circ}50'N,6^{\circ}10'E$), 42 km off the South coast of France.
The basic sensor element of the detector is the Optical Module (OM) 
housing a 10-inch photo-multiplier tube (PMT) inside a pressure resistant glass sphere.
In its final configuration ANTARES has 885 OMs placed along 12 flexible detection lines distributed following an octogonal shape.
The lines are anchored to the seabed by a dead weight placed at the bottom and vertically sustained by means of a buoy at the top. 
Each line is divided in 25 storeys, which are mechanical structures holding a triplet of OMs and the Local Control Module 
(LCM) electronic container.
The OMs are placed symmetrically around the storey vertical axis and facing downward at $45^{\circ}$ with respect to the horizontal plane 
for an increased efficiency for upgoing muon detection. 
PMT signals are discriminated at 1/3 of a photoelectron and digitized inside the LCM by a dedicated ASIC, the Analogue Ring Sampler (ARS), providing their 
amplitude, arrival time and shape. Two ARSs in each PMT are read out in interleaved mode in order
to reduce the electronic dead time. A schematic view of the detector is shown in Figure \ref{fig_antares}.

\section{Time Calibration}
\label{tcalib}
Precise measurement of hit arrival times is crucial for good muon reconstruction. In particular, the absolute time calibration 
is needed to correlate events with astrophysical transient phenomena, while an accurate relative timing is required to achieve
the best angular resolution attainable.
The main uncertainties concerning the absolute calibration come from the electronic paths between the shore station and the detector
lines, while the required event-time accuracy is around 1ms. 
The relative timing refers to the offsets in the arrival photon times between PMTs and is caused by the instrinsic differences among the sensors.
The uncertainties contributing to the spread of the relative offsets (equation \ref{eq_tcont}) come from the transit time spread of the PMT ($\sigma_{TTS}\sim1.3ns$), and the 
effect of the sea water chromatic dispersion in the light propagation ($\sigma_{water}\sim1.5ns$, for a travelled distance of 40 m). 
The contribution due to the front-end electronics can be controlled with a proper calibration and is required to be $\sigma_{elec}\sim0.5$ns.
The different methods and systems used for time calibration purposes in ANTARES \cite{bib_timing} are described in the following sections.

\begin{equation}
\sigma^{2}_{OM} = \frac{\sigma^{2}_{TTS}}{N_{pe}}+\frac{\sigma^{2}_{water}}{N_{\gamma}}+\sigma^{2}_{elec} 
\label{eq_tcont}
\end{equation}

\subsection{The clock system}
A 20MHz clock signal generator on shore is used to provide a common reference signal to the whole apparatus.
The system works by sending optical signals from shore, and through a distribution network, to each LCM  
where a transceiver board sends the signals back as soon as they arrive.
The corresponding round-trip time is, therefore, twice the propagation time along the cables to each individual LCM.\\ 
Measurements of the time delays due to the electronic paths are performed every hour.
Results from the monitoring of the round-trip times between the bottom of a line and one particular storey shown that accuracies of about 16 ps (RMS) are achieved.
The variations observed in the round-trip times of clock signal sent from shore to a certain line are of the order of a few hundred of picoseconds.
The absolute time-stamping of the events is made by interfacing the clock with an electronic card receiving the GPS time, providing a precision 
of the order of 100 ns.

\subsection{Calibration in the dark-room}
Prior to the deployment of a line a complete calibration is performed on shore using a special setup consisting
of a laser sending light trough optical fibre to the OMs placed in a dedicated dark room.
Measuring the difference between the emission time of the laser light and the time when the signal is registered by the PMT the relative time offsets
can be obtained. Afterwards, and taking one OM as a common reference, the individual time differences are corrected and the first calibration constants obtained for 
all the OMs. Those calibration parameters are stored in the experiment's database to be checked and updated (if needed) by the \emph{in-situ} 
measurements.

\subsection{Optical Beacons in-situ calibration}
\begin{figure}
\begin{center}
\includegraphics[width=0.6\linewidth]{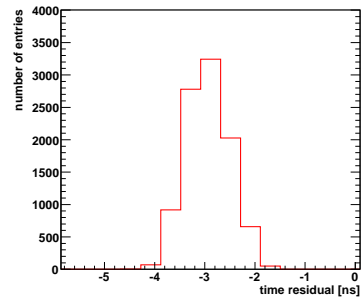}
\caption{Time residuals for an OM located close to the LED beacon. The sigma of the distribution (0.5 ns) gives an estimation of the time resolution due to the electronics.}
\label{fig_close_om}
\end{center}
\end{figure}

The time calibration is performed \emph{in-situ} by an Optical Beacon (OB) system \cite{bib_obeacons}, which consists of LED and laser short pulse light devices 
with a well know emission times.
There are four LED optical beacons (LED OB) placed regulary along each detector line (in storeys 2, 9, 15 and 21). Each LED OB has 36 blue ($\lambda=472$ nm) 
LEDs arranged in groups of six on six vertical boards forming a hexagonal prism. These devices are primarily used to measure the relative time 
differences among OMs and to study the optical properties of the sea water.
Two laser beacons (LOB) emitting green light ($\lambda=532$ nm) are located at the bottom of two central lines. The LOBs are designed 
to illuminate those OMs which can not be reached by the LED beacons and to measure the relative time offsets between lines. In both cases the time offset computation 
is based on the measurement of the time residuals defined as the difference between the light emission time and the
time when the flash is recorded by the PMT.
As an example, Figure \ref{fig_close_om} shows the distribution of the time residuals for an OM placed at a short distance from the LED beacon. 
Because of the high amount of light received by the PMT the contribution from the TTS and the photon dispersion are negligible. Therefore, the sigma ($\sim0.5$ ns) of the distribution gives
directly an estimation of the time resolution due to the electronics, which is well within the specifications.\\

\begin{figure}
\begin{center}
\includegraphics[width=0.6\linewidth]{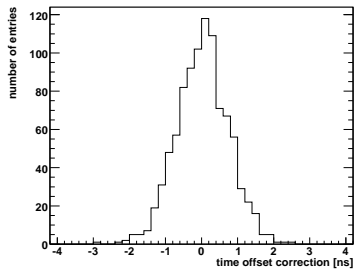}
\caption{Distribution of the corrections measured \emph{in-situ} by the LED OB system to the time offsets obtained on shore.}
\label{fig_insitu}
\end{center}
\end{figure}

The measurement of the relative time offsets between OMs is based on the study of the position of the time residual peak, obtained from a Gaussian fit to the 
region including the rising edge of the time distribution and the first bin after its maximum, as a function of the distance from the OMs to the LED optical beacon. 
It happens that the time residual peak increases with the distance due to the ``early photon'' effect, arising from the 
duration of the light pulse and the fact
that the first photons recorded by the PMT determine the time of the light flash. A specific simulation shows that a linear increase is expected.
The time offset corrections are then obtained as the deviations from a straight line fit to the time residual peaks ordered by distance.
The distribution of these corrections for all the OMs that can be calibrated with the OBs is shown in Figure \ref{fig_insitu}. It is found
that 15$\%$ of the cases need a correction to the dark room offsets larger than 1 ns.\\
The relative time differences among the detector lines are calculated following a similar procedure but using a laser beacon. The results obtained
are compatible with those given by an indepent method based on the time residuals of the reconstructed muon tracks. 

\subsection{Potassium-40 decay}
\begin{figure}
\begin{center}
\includegraphics[width=0.6\linewidth]{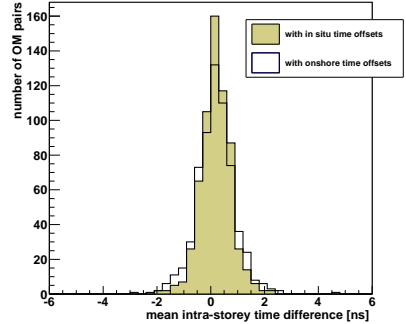}
\caption{Distribution of the intra-storey time differences from the $^{40}$K using the on shore (empty histogram) or the 
values measured \emph{in-situ} (filled histogram).}
\label{fig_k40}
\end{center}
\end{figure}

The $^{40}$K is a radiactive isotope naturally present in the sea water. Its decay can produce an electron with energy 
sufficient to exceed the Cherenkov threshold and induce the light emission.
If this emission occurs in the vicinity of a detector storey a coincidence signal may be recorded by two OMs. Such an event will result in
a visible bump over the distribution of the relative time difference measured by two PMTs in the same storey (intra-storey offsets). If the time offsets of the OMs 
are correct the resulting peak must be centered at zero position.
Figure \ref{fig_k40} shows the distribution of the intra-storey offsets obtained with the $^{40}$K when using the on shore calibration parameters or the 
values measured \emph{in-situ} by the LED OB system. The RMS of the mean intra-storey time difference distribution improves from 0.72 ns to 0.57 ns when using the 
time offsets calculated in situ.

\section{Positioning system}
\label{pcalib}
\begin{figure}
\begin{center}
\includegraphics[width=0.8\linewidth]{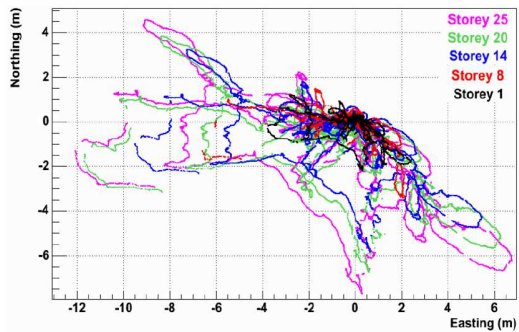}
\caption{Displacements, in the horizontal plane, of the five hydrophones within a certain line as determined by the acousting positioning system.}
\label{fig_3}
\end{center}
\end{figure}

The detector positioning \cite{bib_pos_calib} is obtained from the measurements of two independent systems:
\begin{enumerate}
\item A High Frequency Long Baseline acoustic system (HFLBL) consisting of acoustic emitters and receivers along each line.
\item A tiltmeter and a compass sensor in every storey.
\end{enumerate}
The acoustic system works by measuring the travel times of 40-60 KHz acoustic signals sent by emitting transducers placed at the
anchor of each line, and received by a set of hydrophones regulary placed along a line (one every five storeys). From these measurements, the 
positions of the hydrophones are obtained on the basis of the triangulation principle and a least-mean-square minimization procedure (Figure \ref{fig_3}).
The tiltmeter-compass sensors provide the local tilt angles of each storey with respect the horizontal plane (pitch and roll) as well as 
its orientation with respect the Earth Magnetic North (heading). Using the information gathered with these two systems 
the shape of the lines is reconstructed every two minutes by performing a global $\chi^{2}$ fit based
on a model which predicts the mechanical behaviour of the line under the influence of the sea current.
The relative positions of the OMs are then deduced from the reconstructed line shape and from the known 
geometry of the storeys. The precision achieved with this system is of the order of 10 cm (RMS) even for sea water
currents as strong as 15 cm/s. This corresponds to an uncertainty in the travel time of
light of 0.5 ns that well matches the requirements for track reconstruction.\\

It is also important to know precisely the absolute orientation of the detector in order to discover a cosmic source of neutrinos.
An indirect determination was obtained by sending low frequency acoustic signals from a boat (placed above the ANTARES sea surface) to the acoustic receivers and emitters at the 
bottom of the detector lines. Knowing the position of the boat (equipped with a digital GPS) and using the triangulation principle, the latitude (x) and longitude (y)
of the detector were determined with an uncertainty of $\sigma_{x} = 0.13^{\circ}$ and $\sigma_{y} = 0.04^{\circ}$.

\section{Charge Calibration}
\label{ccalib}
\begin{figure}
\begin{center}
\includegraphics[width=0.7\linewidth]{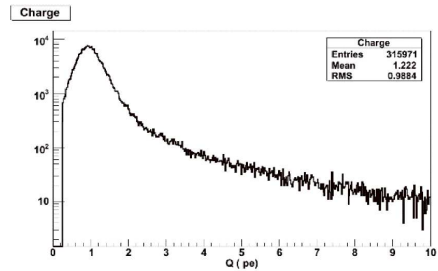}
\caption{Charge distribution calibrated for all the PMTs in the detector.}
\label{fig_charge1}
\end{center}
\end{figure}

\begin{figure}
\begin{center}
\includegraphics[width=0.7\linewidth]{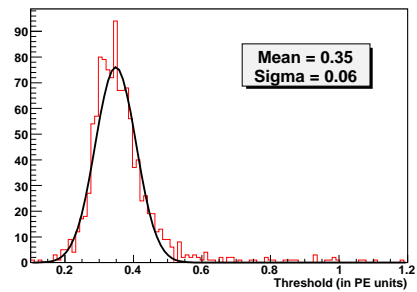}
\caption{Distribution of the effective thresholds after the \emph{in-situ} procedure.}
\label{fig_charge2}
\end{center}
\end{figure}

The charge calibration \cite{bib_charge_calib} enables to translate the signal amplitudes into number of photo-electrons (p.e.), which is the relevant information for muon energy 
reconstruction. 
The charge conversion over the full dynamical range of the ADC can be estimated from the known
position of the single photo-electron peak and of the pedestal. The pedestal value of the AVC channel is measured using special runs during which
the PMT current is digitized at random times.
In the sea water, optical activity from bioluminiscent bacteria or from $^{40}$K decays produce, on average, single photons at the photocathode level. Those minimum bias
events are used to study the single photo-electron peak.
It has been observed that time measurements in the TVC channel influence the charge measurements performed inside the ARS (the inverse effect does not apply).
This ``cross talk effect'' can be corrected on an event-by-event basis using \emph{in-situ} measurements of the AVC value versus the TVC value. 
The maximal size of this correction observed amounts to 0.2  photo-electrons.
Once the cross-talk correction is made, the charge calibration is applied to reconstruct the amplitude of the idividual PMT signals.
The distribution for optical activity events is then single photo-electron charges
as is shown in Figure \ref{fig_charge1}.\\ 
The study of the $^{40}K$ counting rate evolution in time shows a regular decrease which is thought to be caused by the 
ageing of the PMT photocathode. Since all channels are tuned to have an effective threshold of 0.3 p.e. periodic checks need to be performed
and corrections applied when a change of the PMT gain is observed.
This is done using events with a null time stamp (TS=0), which correspond to those cases when the 
amplitude of the signal registered by the PMT is just above the threshold. 
Figure \ref{fig_charge2} shows the effective thresholds distribution obtained after the tuning procedure. 
The mean and sigma improve from the 0.53 and 0.17 respective values measured before applying the \emph{in-situ} corrections.

\section{Conclusions}
\label{conclusions}
The different systems and methods used for calibration purposes in the ANTARES neutrino telescope
have been presented. Measurements done at the integration sites provided
the first calibration parameters. Further checks performed \emph{in-situ} confirm the high level of precision
achieved. The adopted systems ensure an optimal performance of the telescope. The detector calibration is under control and the experiment is routinely taking data.

\section{Acknowledgements}
We gratefully acknowledge the financial support of the Spanish
Ministerio de Ciencia e Innovaci\'on (MICINN), grants
FPA2009-13983-C02-01, ACI2009-1020 and Consolider MultiDark
CSD2009-00064 and of the Generalitat Valenciana, Prometeo/2009/026. 








\end{document}